\documentclass[12pt,letterpaper]{iopart}
\usepackage [dvips]{graphics, color}
\usepackage{iopams,wrapfig}%for IOP style package

\begin{document}

%short title/ long title
\article[System size dependence of strange particle correlations]{Strangeness in Quark Matter}{System size dependence of strange particle correlations in Cu+Cu and Au+Au collisions at $\sqrt{s_{NN}}$ = 200 GeV in STAR at RHIC}
\author{C E Nattrass for the STAR Collaboration}
\address{WNSL, 272 Whitney Ave., Yale University, New Haven, CT 06520, USA}
\ead{christine.nattrass@yale.edu}

%\maketitle

\begin{abstract}
%Should not normally exceed 200 words.
Two particle azimuthal correlations with strange trigger particles ( $K^0_S,\Lambda,\Xi$ ) are presented in Cu+Cu and Au+Au collisions at $\sqrt{s_{NN}}$ = 200 GeV in STAR at RHIC.  The near-side associated yield is investigated as a function of centrality, $p_T$, and strangeness content of the trigger to investigate possible flavour or meson/baryon differences.  The system size dependence of both the jet-like correlations and the long range pseudorapidity correlations is studied to give insight into the origin of these two components.
\end{abstract}

\section{Introduction}

 \begin{wrapfigure}{r}{8cm}
%\vspace{0.1cm}
%\hspace{0.1cm}
%\begin{figure}
%Thr * after figure makes it span both columns
\vspace{-.75cm}
\begin{center}
%\mbox{\resizebox{6cm}{!}{\includegraphics{AssocPt.eps}}}
\rotatebox{0}{\resizebox{8cm}{!}{\includegraphics{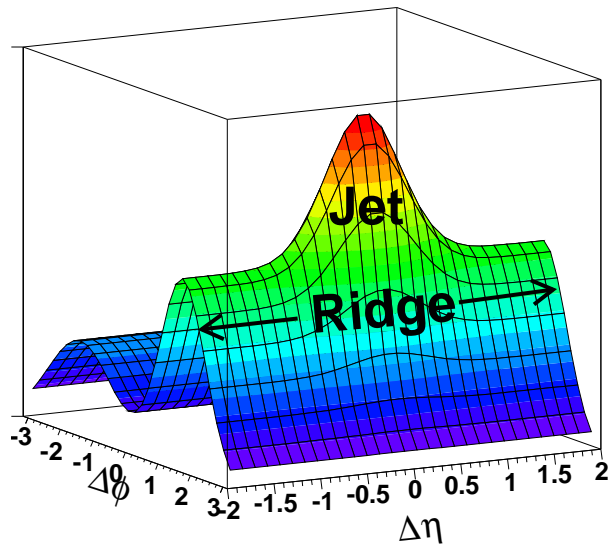}}}
\label{NearSide}
%\begin{flushleft}
\caption{A schematic diagram demonstrating the structure of the near-side jet.  Colour online.}
%\end{flushleft}
%\vspace{0.1cm}
\vspace{-.75cm}
\end{center}
\end{wrapfigure}
Previous studies at RHIC have demonstrated suppression of inclusive spectra in A+A collisions relative to p+p \cite{phenixRAA,starRAA}, along with enhanced baryon/meson ratios in the intermediate $p_T$ range 2-6 GeV/c \cite{starLK,starLK2}.  These results indicate that unmodified jet fragmentation is not the dominant production mechanism in this $p_T$ range.  Additionally, meson and baryon spectra follow different trends in the intermediate $p_T$ region, making this an interesting regime to study identified particles in jets using two-particle correlations.

Indeed, novel features have been observed in the near-side jet peak at intermediate $p_T$ in heavy ion collisions \cite{Joern}.  These features are drawn schematically in \fref{NearSide}.  A peak which is narrow in both $\Delta\phi$ and $\Delta\eta$ similar to jets in elementary collisions, called the \textit{Jet}, is present but sits on top of a structure called the \textit{Ridge} which is also narrow in $\Delta\phi$ but broad in $\Delta\eta$.

Studies of both structures in two-particle correlations using strange particle triggers ($K^0_S, \Lambda, \Xi$) with associated charged particles (h) are presented here for Cu+Cu collisions at $\sqrt{s_{NN}}$ = 200 GeV and compared to Au+Au collisions at the same energy in STAR.  Details of the STAR detector can be found in \cite{STARNIM}.  The near-side is studied as a function of system size, centrality, $p_T$, and strangeness content in order to look for flavour, baryon/meson, and system size dependencies.

\section{Data Analysis}

\begin{figure}
%Thr * after figure makes it span both columns
%These figures - I need to put this in in log scale and the d+Au on the same figure is nice and I need to add (a) and (b) to the figures.
\begin{center}
\hspace{0.0cm} 
\rotatebox{0}{\resizebox{15.7cm}{!}{\includegraphics{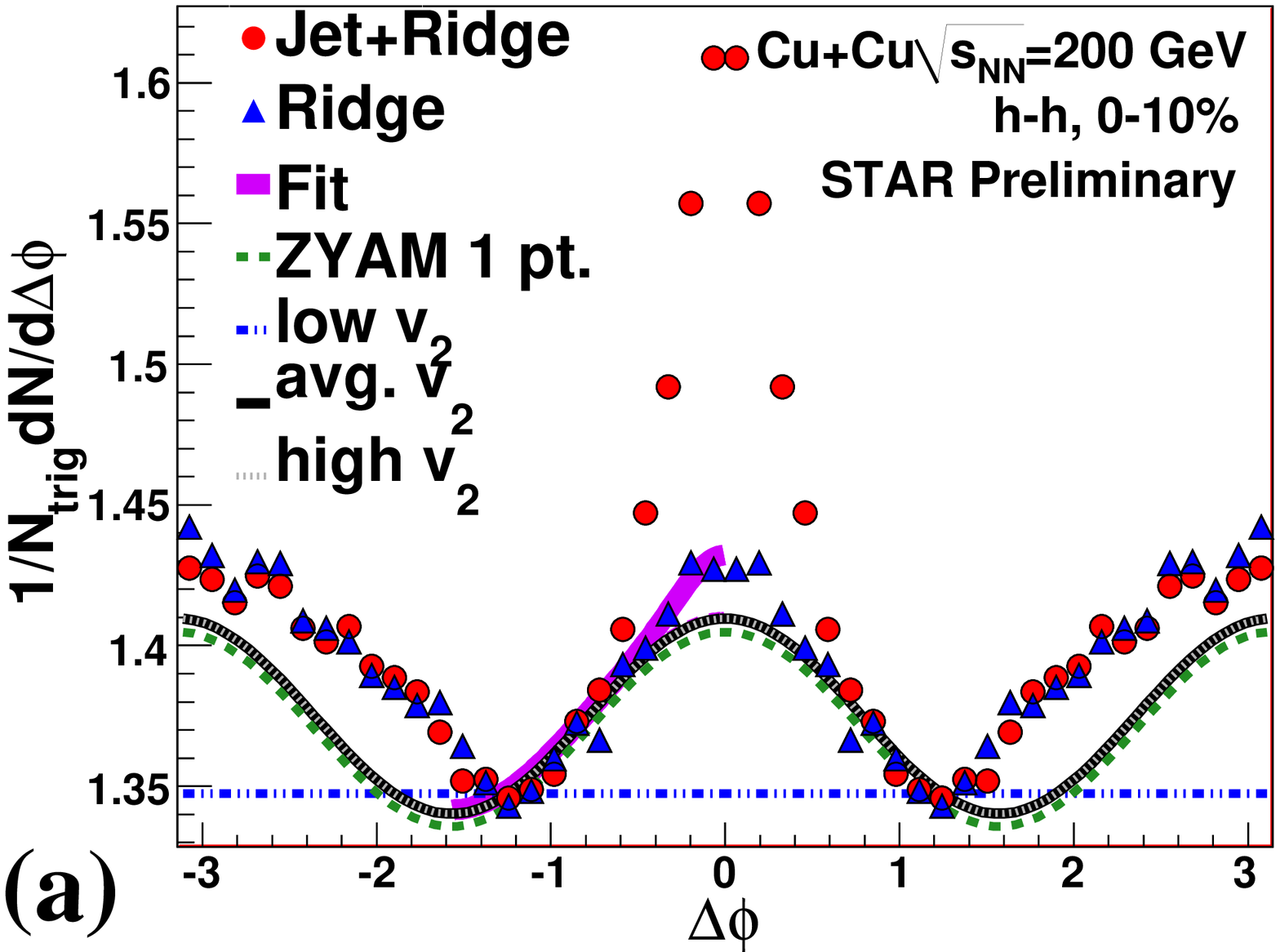}\includegraphics{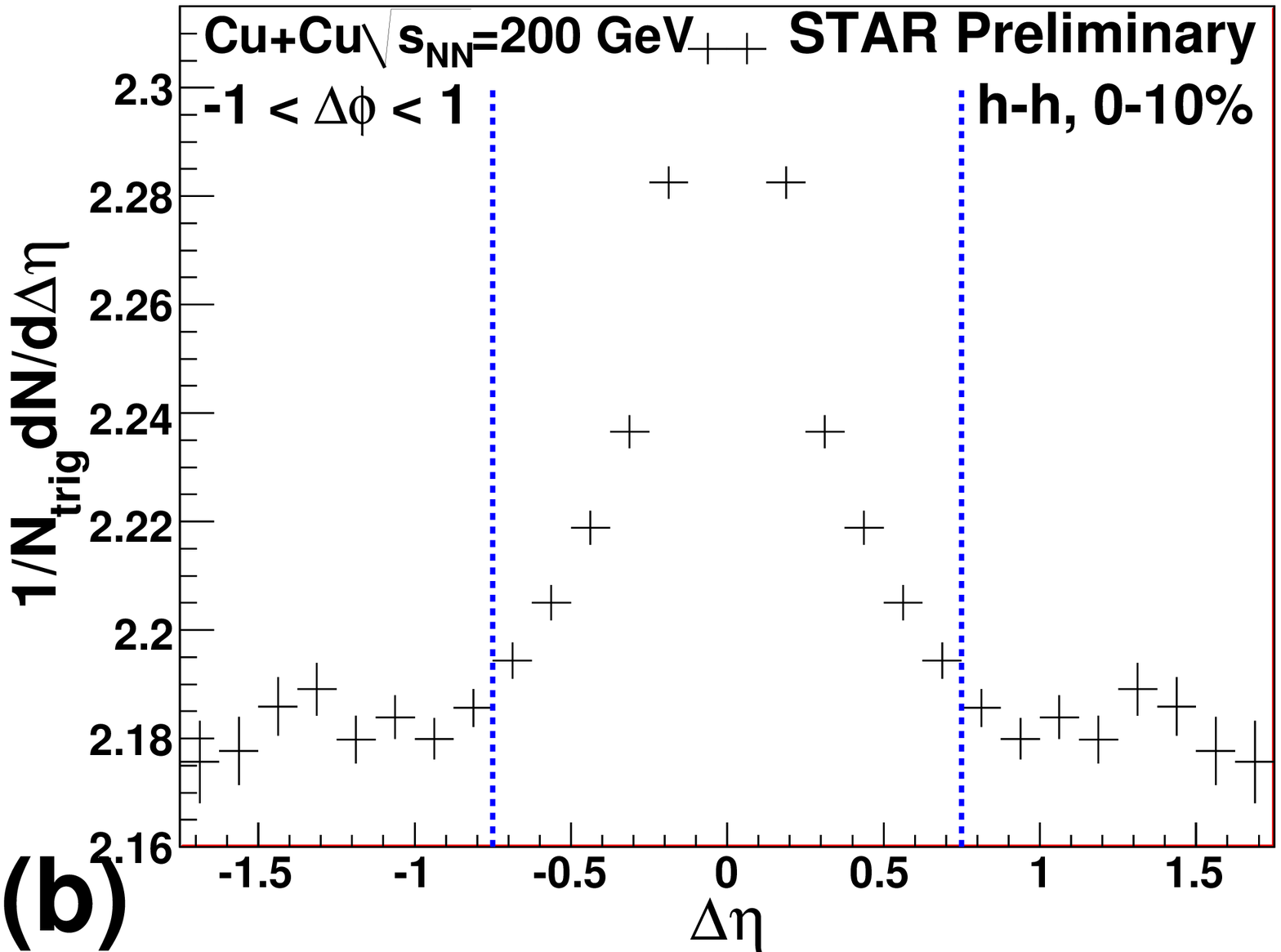}}}
%\rotatebox{0}{\resizebox{15.7cm}{!}{\includegraphics{EtaProj.eps}}}
\caption{(a) Azimuthal distribution of particles in the \textit{Jet+Ridge} ($|\Delta\eta| < 0.75)$ and \textit{Ridge} ($0.75< |\Delta\eta| < 1.75$) regions with background determined for the \textit{Ridge} histogram as described in text.  The \textit{Ridge} histogram has been scaled by 0.75 to match the $\Delta\eta$ window of the \textit{Jet+Ridge} histogram.  The backgrounds from different methods are so close they cannot be distinguished.  The background with the average $v_2$ is determined using the ZYAM method with three points. (b) $\Delta\eta$ projection of the distribution of particles in the range $|\Delta\phi|<1$.  Vertical lines indicate where the cuts for the \textit{Jet} and the \textit{Ridge} regions are made.  Data shown are for $3.0$ GeV/c $< p_{T}^{trigger} < 6.0 $ GeV/c and $1.5$ GeV/c $< p_{T}^{associated} < p_{T}^{trigger}$ h-h correlations in Cu+Cu collisions at $\sqrt{s_{NN}}$ = 200 GeV.  Colour online.}
\vspace{0.0cm} 
\end{center}
\label{RidgeProjections}
\end{figure}

The azimuthal distributions of pairs of particles normalized by the number of trigger particles are corrected for the single particle reconstruction efficiency of associated particles, elliptic flow ($v_2$), azimuthal acceptance of the detector, and the acceptance in $\Delta\eta$.  A sample of such a distribution of particles function is shown in \fref{RidgeProjections}(a).  The y-axis is the number of particles associated with a single trigger particle per unit $\Delta\phi$ so that the integral after background subtraction is the number of particles associated with the jet.  Clear near- and away-side jets are visible above the background, indicating that at least some of these hadrons originate from di-jets.  The $p_T$ of the both the trigger particle and the associated particles are restricted to limit the background.  The near-side yield of associated particles is determined by fitting a Gaussian to the near-side peak.  The \textit{Jet} and \textit{Ridge} contributions to the near-side peak were separated by studying the distribution of particles in different $\Delta\eta$ windows.  The \textit{Ridge} was previously observed to be approximately flat in $\Delta\eta$ in Au+Au collisions\cite{Joern}, so to determine the \textit{Jet} yield the normalized distribution of particles in the \textit{Ridge} yield was subtracted from the that in the \textit{Jet} region.  To maximize the statistics, the \textit{Ridge} yield was determined in the $0.75 < |\Delta\eta| < 1.75$ region and therefore was scaled by the ratio of the widths of the \textit{Jet} and \textit{Ridge} regions in $\Delta\eta$ (0.75) in order to subtract the \textit{Ridge} contribution in the range $|\Delta\eta| < 0.75$.  Assuming that $v_2$ is flat in $\eta$, a reasonable assumption in the mid-rapidity range $|\eta| < 1$ based on the available data \cite{phobosFlow1,phobosFlow2}, any systematic error due to flow subtraction will be constant across $\Delta\eta$ and therefore cancel out when determining the \textit{Jet} yield using this method.  The \textit{Jet} yield was then determined by fitting to a Gaussian plus a constant.

%\begin{wrapfigure}{r}{8cm}
%\end{wrapfigure}

For the \textit{Ridge}, systematic errors due to the error on $v_2$ and errors in the level of the combinatorial background are determined by comparing different methods.  \Fref{RidgeProjections}(a) shows an example of an azimuthal distribution of particles before background subtraction along with the various estimates of the background determined as described below.  The systematic errors from the level of the combinatorial background are determined by comparing the Zero Yield At Minimum (ZYAM) method \cite{phenixZYAM} using one point, the ZYAM method averaging over three points, and allowing the background level to vary as a free parameter.  The bounds on $v_2$ were determined by comparing different methods for the $v_2$ measurement.  $v_2\{FTPC\}$ is from the event plane method using tracks found in the Forward Time Projection Chamber (FTPC).  $v_2\{CuCu-pp\}$ is the flow measured via azimuthal correlations with non-flow contributions,  as measured in p+p collisions, subtracted.  The higher of the two is used to determine the upper bound on $v_2$.  The lower bound on the $v_2$ is either 0 for the 0-10\% centrality bin or $\sqrt{v_{2,centrality}^2-v_{2,0-10\%}^2\frac{M_{centrality}}{M_{0-10\%}}}$ where M is the mean reference multiplicity, for the other bins.  The nominal value for the yield is determined using the background level from the fit and the average of $v_2\{FTPC\}$ and $v_2\{CuCu-pp\}$.  The systematic error due to $v_2$ is the dominant systematic error for most data points.

Corrections for losses due to track merging and track crossing, affecting tracks close in both $\Delta\phi$ and $\Delta\eta$, have not been done yet.  The depletion is a function of the overall track density and the relative effect is therefore larger at low $p_T$ where the combinatorial background is large.  Track merging is also greater for the $\Lambda$ and $K^0_S$ and greatest for the $\Xi$ because each daughter track can merge with the associated tracks.  More details can be found in \cite{Marek}.  With the method used for separating the \textit{Jet} and the \textit{Ridge}, the effect should be greater in the \textit{Jet} than in the \textit{Ridge} because it affects tracks which are close in both azimuth and pseudorapidity, however, this has not been studied quantitatively.  In the kinematic range studied, track merging effects are estimated to be less than 15\%.  Therefore, a 15\% systematic error has been added to the statistical error.

\section{Results}
To illustrate the ridge effect in Cu+Cu collisions, \fref{RidgeProjections}(b) shows that the \textit{Jet} does not extend beyond the $|\Delta\eta|<0.75$, and \fref{RidgeProjections}(a) shows that both $\Delta\eta$ ranges have yield above the background level, even given the errors in $v_2$ and the background level.  This demonstrates that the data are consistent with the presence of the \textit{Ridge} in Cu+Cu.

\begin{figure}[t!]
\rotatebox{0}{\resizebox{15.7cm}{!}{\includegraphics{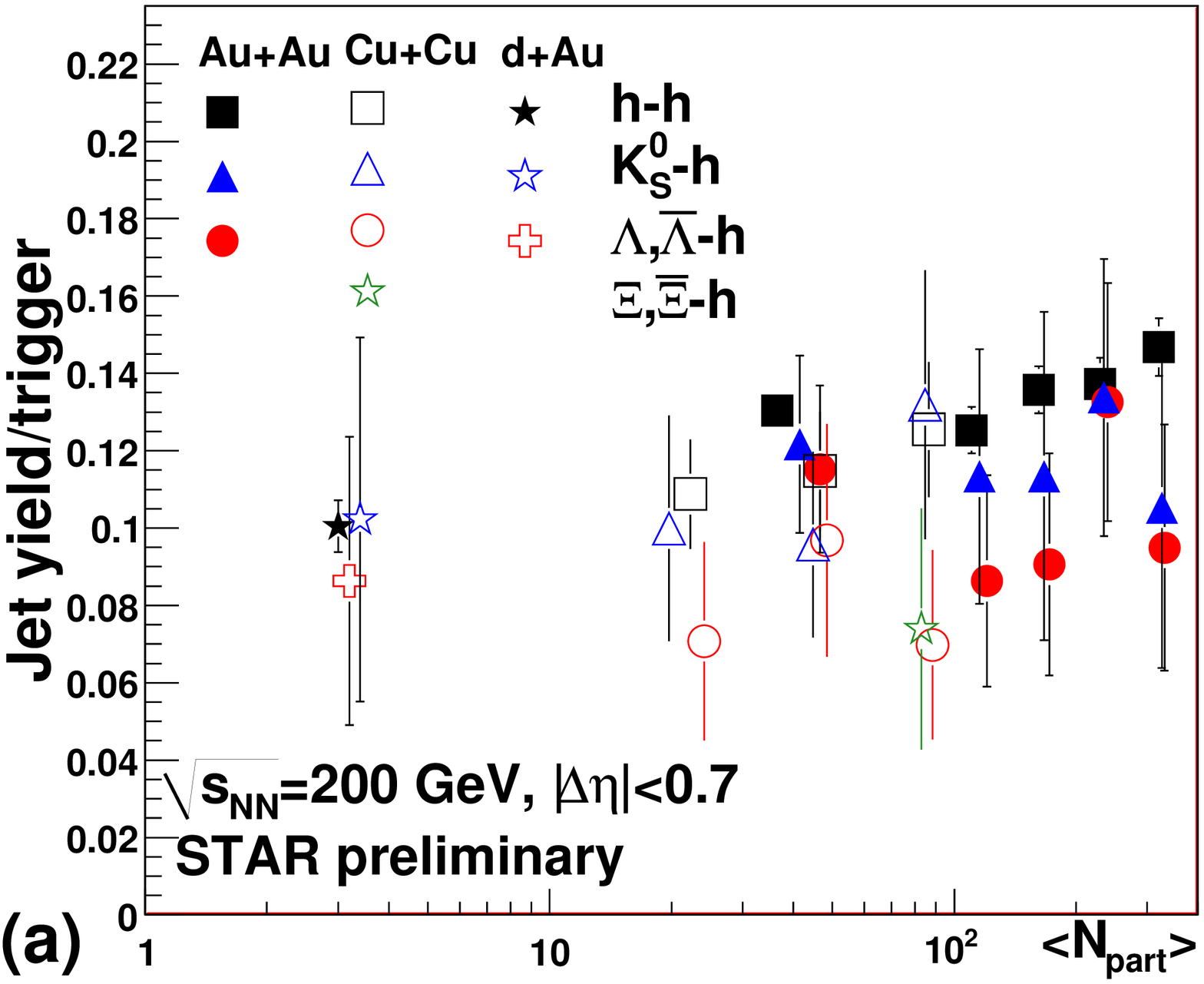}\includegraphics{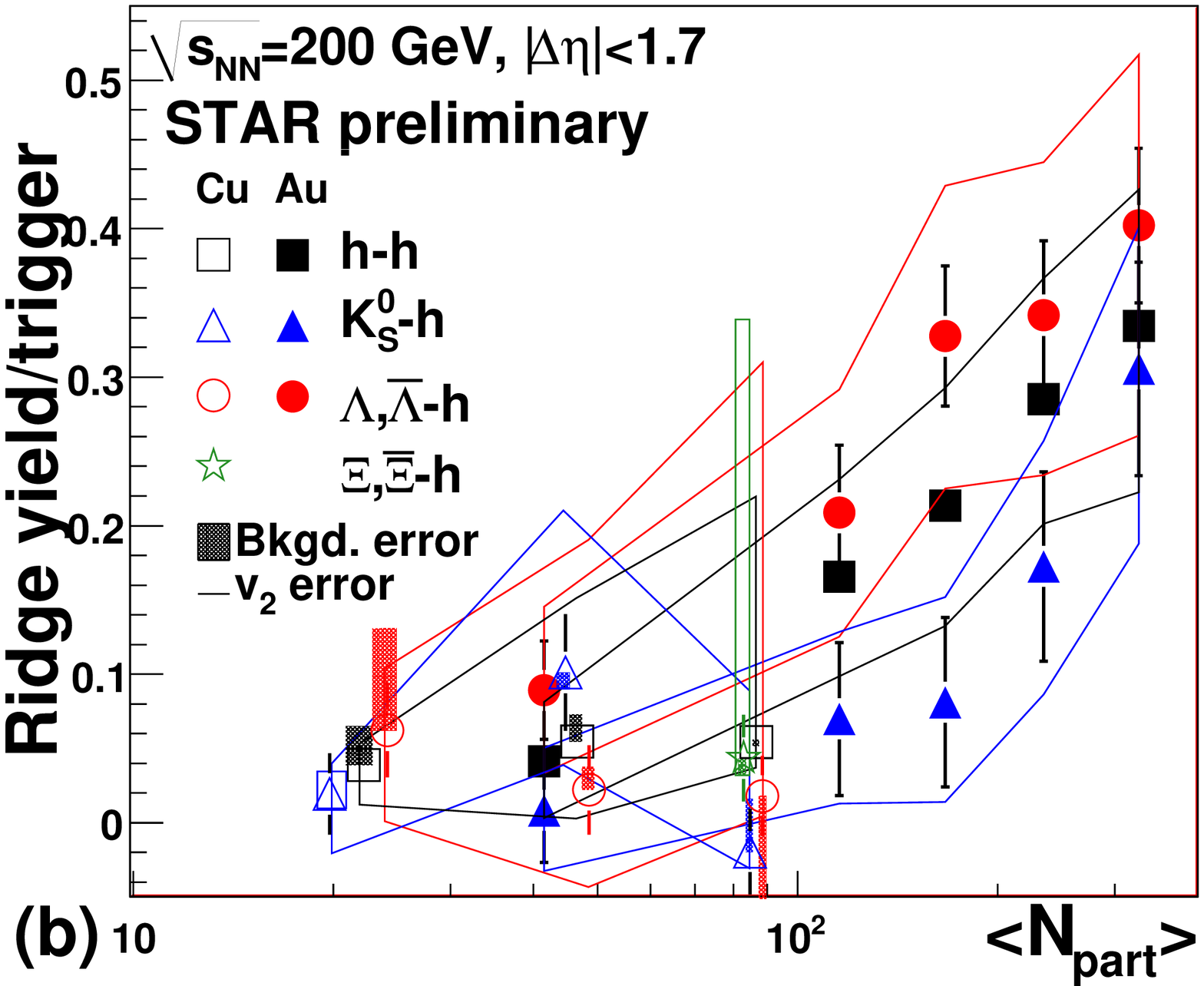}}}
\caption{(a) \textit{Jet} yield per trigger as a function of $<N_{part}>$. (b) \textit{Ridge} yield per trigger as a function of $<N_{part}>$.  Au+Au and d+Au data from \cite{Jana}.  As a result of the smaller system size in Cu+Cu, these data do not extend as far in $<N_{part}>$ as the Au+Au data.  In both figures 3.0 GeV/c $< p_T^{trigger} <$ 6.0 GeV/c and 1.5 GeV/c $< p_T^{associated} < p_T^{trigger}$.  Colour online.}
\label{JetAndRidgeNPart}
\end{figure}

The \textit{Ridge} and \textit{Jet} yields as a function of $<N_{part}>$ are shown in \fref{JetAndRidgeNPart}.  This indicates that the jet-like part of the near-side peak, which is narrow in both azimuth and pseudorapidity like the jet peak observed in vacuum fragmentation, is independent of system size, and trigger species.  No significant dependence of the jet-like yield on centrality is observed.  This implies that the jet-like correlation may be dominated by unmodified jets.

The \textit{Ridge} yields in Cu+Cu shown in \fref{JetAndRidgeNPart}(b) are consistent with the Au+Au \textit{Ridge} yields at the same $<N_{part}>$.  The systematic errors on the Cu+Cu data inhibit the inference of a trend on the basis of the Cu+Cu data alone.  However, Au+Au and Cu+Cu data combined exhibit a monotonically increasing \textit{Ridge} yield as a function of $<N_{part}>$, and there is no strong dependence on the collision system.  There may be a dependence on the particle type, which could be confirmed by future studies if the systematic error due to $v_2$ subtraction can be decreased.

\begin{figure}
%Thr * after figure makes it span both columns
%\rotatebox{0}{\resizebox{15.7cm}{!}{\includegraphics{JetYieldPtCuCuCent.eps}\includegraphics{JetYieldPtCuCuIDPart.eps}\includegraphics{JetYieldPtAuAuIDPart.eps}}}\caption{Caption}
\rotatebox{0}{\resizebox{15.7cm}{!}{\includegraphics{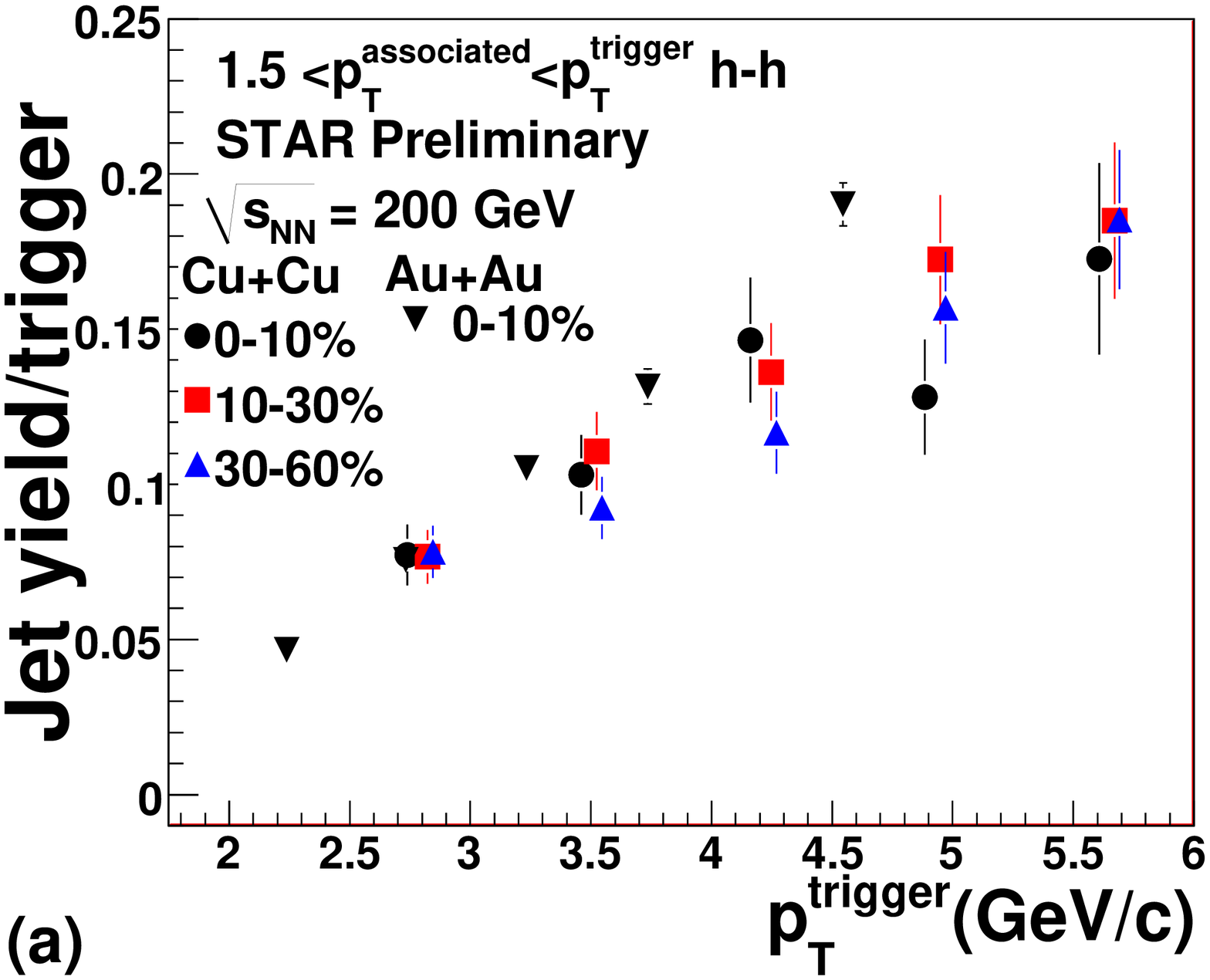}\includegraphics{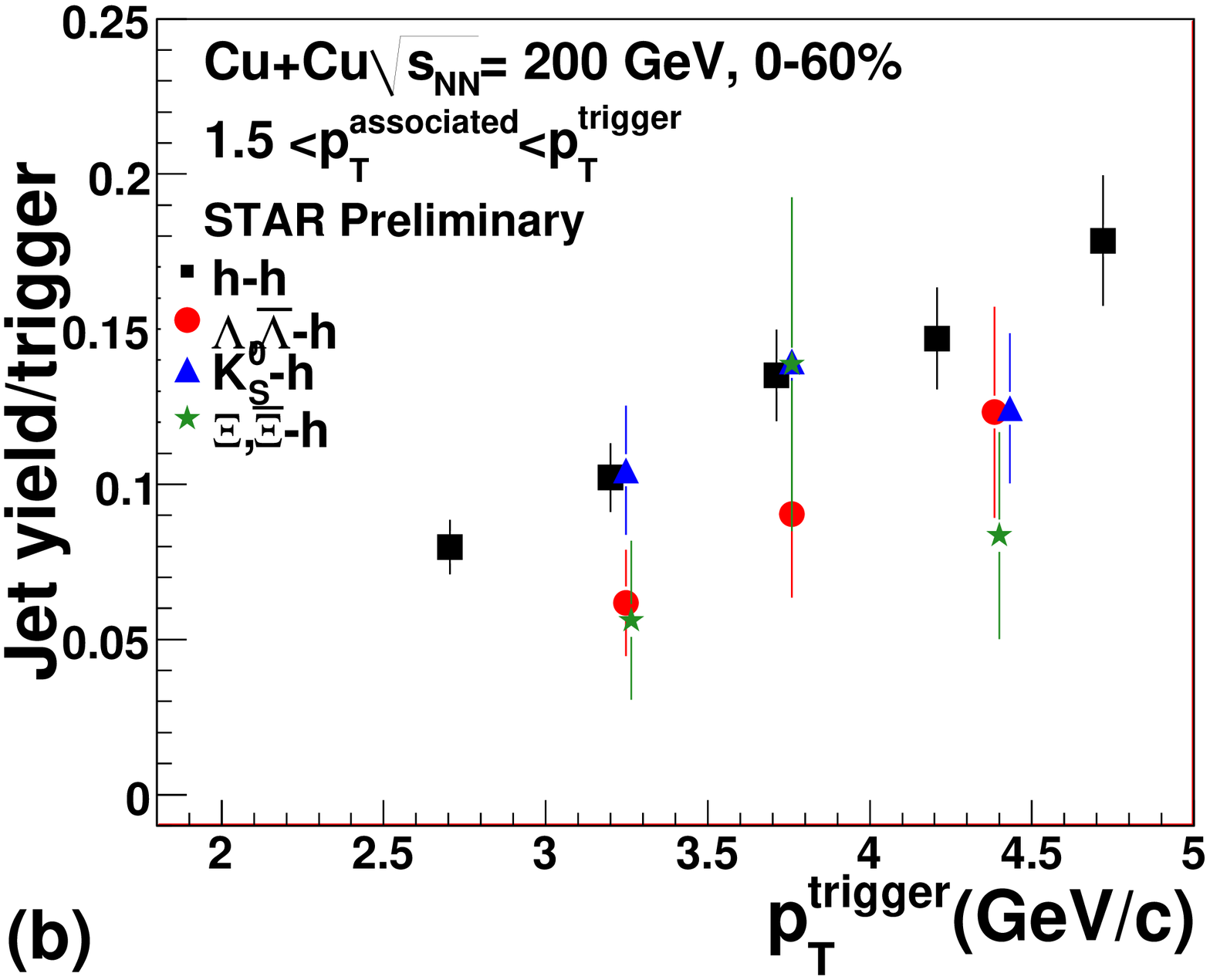}}}\caption{(a) h-h near-side yield per trigger for different centralities in Cu+Cu collisions and 0-10\% Au+Au collisions as a function of $p_{T}^{trigger}$.  10-30\% Cu+Cu data are offset by 30 MeV, 30-60\% data are offset by 60 MeV, and 0-10\% Au+Au data were offset by 90 MeV for visibility.  (b) Near-side yield for different trigger particles as a function of $p_{T}^{trigger}$.  $\Lambda$ data were offset by 30 MeV, $K^0_S$ by 60 MeV, and $\Xi$ 90 MeV for visibility.  Colour online.}
\label{JetPt}
\end{figure}

\Fref{JetPt}(a) shows the h-h \textit{Jet} yield per trigger as a function of $p_{T}^{trigger}$ for different centrality bins in Cu+Cu collisions and in the 0-10\% most central Au+Au collisions.  The yield monotonically increases as a function of $p_{T}^{trigger}$ and shows no dependence on centrality or system size within errors.  Since there is no observed centrality dependence, the 0-60\% Cu+Cu data are combined to 
improve statistics for a comparison of trigger particle species, shown in \fref{JetPt}(b).

As observed in Au+Au \cite{Jana}, the identified particle triggered \textit{Jet} yields tend to be slightly lower than the unidentified trigger particles in both \fref{JetAndRidgeNPart}(a) and \fref{JetPt}, however, these data have not been corrected for track merging which is known to decrease the apparent yield and to be dependent on the number of tracks used to reconstruct the particle.  Within the errors, there are no particle type differences in the \textit{Jet}.

%We have hard-wired the figure number here... not elegant but easier than debugging the latex
Analysis of the associated particle spectra in Au+Au at $\sqrt{s_{NN}}$ = 200 GeV demonstrated that the particles in the jet-like peak had an inverse slope parameter (590 $\pm$ 30 MeV, 4.0 $< p_T^{trigger} <$ 6.0 GeV/c) considerably higher than the inclusive spectrum (355 $\pm$ 6 MeV, fit above 2.0 GeV), whereas the spectra of the associated particles in the \textit{Ridge} (395 $\pm$ 23 MeV, 4.0 $< p_T^{trigger} <$ 6.0 GeV/c) were comparable to the inclusive spectra \cite{Joern,Jana}. The $p_T$ spectra of the associated charged particles in Cu+Cu collisions in the \textit{Jet} region,  determined by fitting the near-side yield in different $p_{T}^{associated}$ bins with 3.0 GeV $< p_T^{trigger} <$ 6.0 GeV, are shown in figure 5.  The inverse slope for the \textit{Jet} unidentified hadron triggers in Cu+Cu collisions from an exponential fit over the data shown is $445 \pm 20$ MeV (statistical errors only, 3.0 $< p_T^{trigger} <$ 6.0 GeV/c), comparable to the inverse slope of $478 \pm 8$ MeV (statistical errors only, 3.0 $< p_T^{trigger} <$ 6.0 GeV/c) observed in the Au+Au in the same kinematic region.  The difference between the inverse slopes in the two trigger $p_T$ ranges is believed to be due to the selection of higher energy jets for a higher trigger $p_T$.

\section{Conclusions}
We have presented a study of the near-side yield associated with $K^0_S$, $\Lambda$, and $\Xi$ trigger\begin{wrapfigure}{r}{8cm}
\vspace{-.3cm}
\begin{center}
\rotatebox{0}{\resizebox{8cm}{!}{\includegraphics{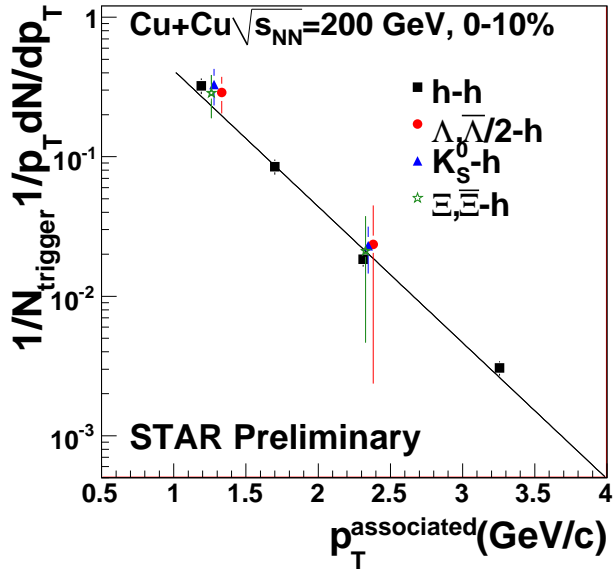}}}
\label{AssocPt}
\hspace{-1cm}
\caption{Near-side \textit{Jet} yield as a function of $p_{T}^{associated}$ in Cu+Cu collisions at $\sqrt{s_{NN}}$ = 200 GeV.  An exponential fit in $p_T$ is shown for h-h.  The inverse slope from the fit was  $445 \pm 20$ MeV (statistical errors only.)  The trigger $p_T$ is restricted to 3.0 GeV/c $< p_T^{trigger} <$ 6.0 GeV/c.  Colour online.}
\end{center}
\vspace{-1cm}
\end{wrapfigure} hadrons and compare with results for unidentified hadrons. We decompose the near-side yield into a short range \textit{Jet} contribution and a long-range (in $\Delta\eta$) \textit{Ridge} contribution.  The jet-like correlation appears to depend on the $p_T$ of the trigger and associated particles, shows no significant dependence on trigger particle type, collision system, or centrality.  The $p_T$ spectra of associated particles in h-h correlations in Cu+Cu collisions in the jet-like correlation are consistent with the $p_T$ spectra observed in jet-like correlations from Au+Au collisions.  The yield per trigger of the long-range pseudorapidity correlations - the \textit{Ridge} - is consistent with an $<N_{part}>$ dependence independent of the collision system and trigger particle type within the errors.  This is consistent with the jet-like peak being dominated by hard parton fragmentation and the long-range pseudorapidity correlations with a hard parton undergoing medium modification.

\end{document}